\documentclass[conference]{IEEEtran}
\IEEEoverridecommandlockouts

\usepackage[minbibnames=1, maxbibnames=6, style=ieee, mincitenames=1, maxcitenames=1, url=false]{biblatex}
\usepackage{amsmath,amssymb,amsfonts}
\usepackage{booktabs}
\usepackage{threeparttable}
\usepackage{algorithmic}
\usepackage{graphicx}
\usepackage{svg}
\svgpath{{./figures/}}

\ifCLASSOPTIONcompsoc
  \usepackage[caption=false,font=normalsize,labelfont=sf,textfont=sf]{subfig}
\else
  \usepackage[caption=false,font=footnotesize]{subfig}
\fi

\usepackage{textcomp}
\usepackage{xcolor}
\usepackage[colorlinks=true,urlcolor=teal,linkcolor=teal,citecolor=teal]{hyperref}

\def\BibTeX{{\rm B\kern-.05em{\sc i\kern-.025em b}\kern-.08em
    T\kern-.1667em\lower.7ex\hbox{E}\kern-.125emX}}

\makeatletter
    \newcommand{\linebreakand}{
      \end{@IEEEauthorhalign}
      \hfill\mbox{}\par
      \mbox{}\hfill\begin{@IEEEauthorhalign}
    }
\makeatother

\begin{document}

\title{
First Experience with Real-Time Control Using Simulated VQC-Based Quantum Policies
\thanks{The project this report is based on was supported with funds from the German Federal Ministry for Economic Affairs and Climate Action in the funding program \emph{Quantum Computing – Applications for industry} under project number 01MQ22008B.
The sole responsibility for the report's contents lies with the authors. This research was carried out within the Munich Center for Machine Learning (MCML), funded by the German and Bavarian AI Strategy.
}
}


\author{
\IEEEauthorblockN{Yize Sun$^*$}
   \IEEEauthorblockA{
       \textit{Siemens AG \& MCML \& LMU}\\
Munich, Germany \\
yize.sun@siemens.com}
\and
\IEEEauthorblockN{Mohamad Hagog$^*$}
   \IEEEauthorblockA{
       \textit{LMU Munich}\\
Munich, Germany \\
m.hgog@campus.lmu.de}
\and
\IEEEauthorblockN{Marc Weber}
   \IEEEauthorblockA{
       \textit{Siemens AG}\\
Munich, Germany \\
marc-weber@siemens.com}
\and
\IEEEauthorblockN{Daniel Hein}
   \IEEEauthorblockA{
       \textit{Siemens AG}\\
Munich, Germany \\
hein.daniel@siemens.com}
\linebreakand 
\IEEEauthorblockN{Steffen Udluft}
   \IEEEauthorblockA{
       \textit{Siemens AG}\\
Munich, Germany \\
steffen.udluft@siemens.com}
\and
\IEEEauthorblockN{Volker Tresp$^\dagger$}
\IEEEauthorblockA{
\textit{LMU Munich \& MCML}\\
Munich, Germany \\
volker.tresp@lmu.de}
\and
\IEEEauthorblockN{Yunpu Ma$^\dagger$}
   \IEEEauthorblockA{
       \textit{LMU Munich \& MCML}\\
Munich, Germany \\
cognitive.yunpu@gmail.com}
}

\maketitle
\def\thefootnote{*}\footnotetext{These authors contributed equally to this work.}
\def\thefootnote{$\dagger$}\footnotetext{Corresponding authors}

\begin{abstract}
This paper investigates the integration of quantum computing into offline reinforcement learning and the deployment of the resulting quantum policy in a real-time control hardware realization of the cart-pole system. 
Variational Quantum Circuits (VQCs) are used to represent the policy. 
Classical model-based offline policy search was applied, in which a pure VQC with trainable input-output weights 
is
used as a policy network instead of a classical multilayer perceptron.
The goal is to evaluate the potential of deploying quantum architectures in real-world industrial control problems. 
The experimental results show that the investigated model-based offline policy search is able to generate quantum policies that can
balance the hardware cart-pole. A latency analysis reveals that while local simulated execution meets real-time requirements, cloud-based quantum processing remains too slow for closed-loop control.

\end{abstract}

\begin{IEEEkeywords}
Quantum Computing, Quantum Reinforcement Learning, Variational Quantum Circuits, Offline Reinforcement Learning
\end{IEEEkeywords}

\section{Introduction}
Quantum computing is a rapidly advancing field leveraging quantum mechanical principles like superposition, entanglement, and interference to perform computations beyond classical limitations.
By operating on quantum bits (qubits), it enables speedups for specific tasks and has shown impact in fields like quantum cryptography~\cite{Bennett_2014}, chemistry~\cite{aspuru2005simulated}, and optimization~\cite{farhi2019quantumsupremacyquantumapproximate}.
    
In the Noisy Intermediate-Scale Quantum (NISQ) era, practical quantum computing remains challenging due to hardware limitations. To address these, hybrid approaches such as Variational Quantum Circuits (VQCs) have gained traction. These circuits combine quantum operations with classical optimization and have been applied to machine learning and reinforcement learning (RL) tasks~\cite{Cerezo_2021}.

RL enables agents to learn decision-making strategies through interaction and has seen success in robotics, healthcare, and industrial automation~\cite{Kober_2013, yu2020reinforcementlearninghealthcaresurvey, LI202375}. Quantum Reinforcement Learning (QRL), which combines RL with quantum methods, has been proposed as a way to improve sample efficiency and optimization~\cite{Dunjko_2016, chen2020variationalquantumcircuitsdeep, kölle2024quantumadvantageactorcriticreinforcement, kolle2024study}.

This paper presents a practical exploration of integrating quantum computing into offline RL for real-time control of a physical cart-pole system. We focus on the MOdel-based Offline policy Search with Ensembles (MOOSE) \cite{swazinna2021overcoming} algorithm—a model-based offline RL method that learns policies using synthetic rollouts from an ensemble of learned dynamics models. In our approach, we replace the classical neural network policy in MOOSE with a purely quantum policy based on a VQC.

To enhance the trainability of the quantum policy, we extend a common circuit design by introducing trainable input and output weights, following the ideas of \cite{skolik2022quantum}. This allows for feature-dependent scaling during encoding and flexible post-processing after measurement without relying on additional classical layers. Unlike hybrid quantum-classical approaches that sandwich a VQC between classical layers, we propose a design in which all trainable parameters are contained within the quantum circuit.

The goal of this work is to investigate whether such a VQC-based policy, trained offline and without classical neural components, can handle a real-world control task under hardware and latency constraints. By deploying the trained quantum policy on a physical cart-pole system, we aim to gain insights into the learning behavior, applicability, and limitations of quantum models in industrially relevant RL settings.

\section{Architecture}

\subsection{Basic Ansatz}
Our policy is implemented as a VQC composed of eight qubits and follows an architecture inspired by the StronglyEntanglingLayers template \cite{Schuld_2020}. The circuit is structured into two layers, each consisting of parametrized single-qubit rotations and CNOT entangling operations arranged in a ring topology.

\begin{figure}[!htbp]
    \centering
    \includegraphics[width=0.95\linewidth]{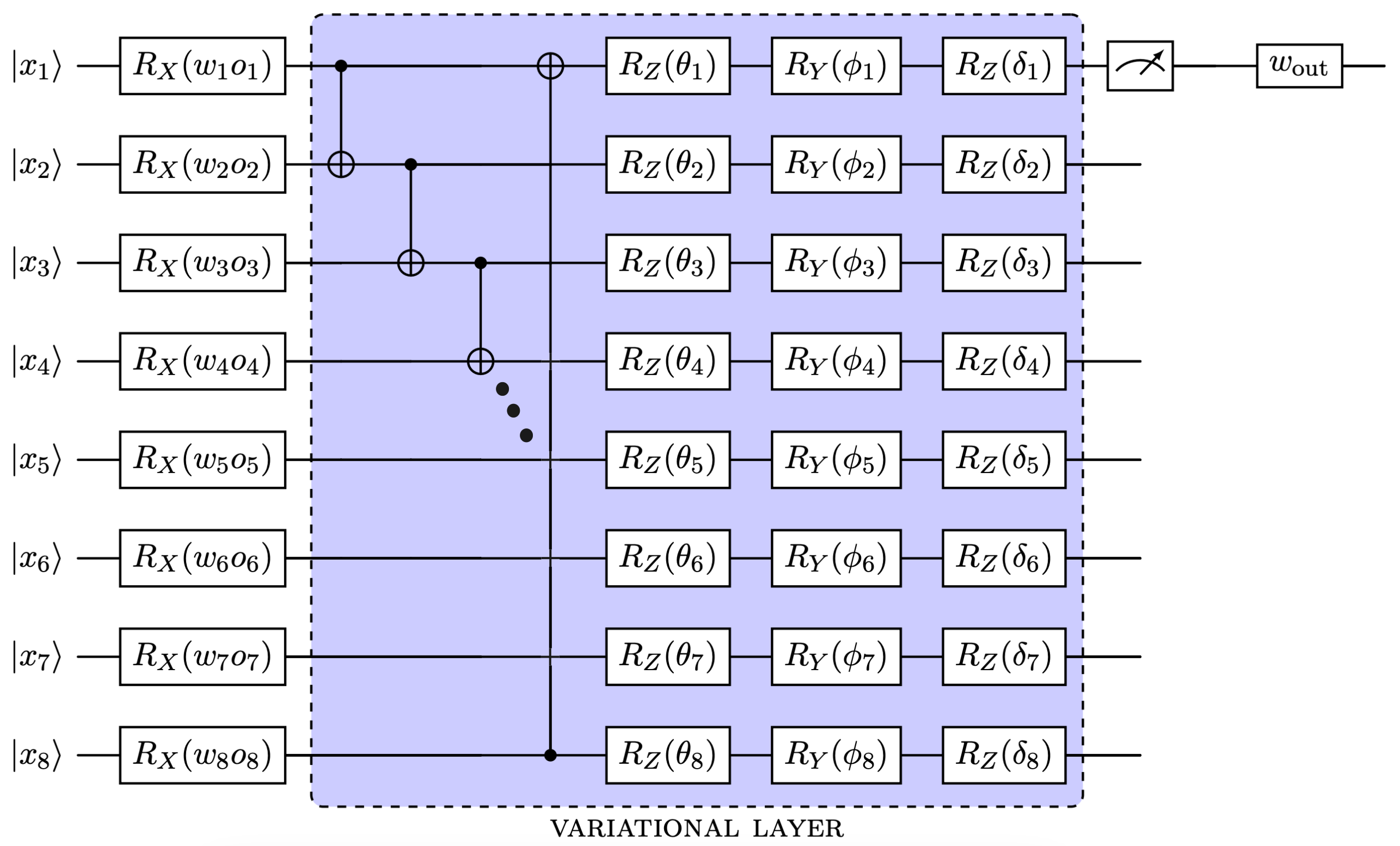}
    \caption{Classical inputs are encoded via trainable $R_X(w_ix_i)$ rotations. Each variational layer consists of $R_Z$–$R_Y$–$R_Z$ rotations and CNOT entanglement. The output is obtained from a computational basis measurement and scaled by a trainable weight $w_\text{out}$.}
    \label{fig:vqc_architecture}
\end{figure}

Figure~\ref{fig:vqc_architecture} illustrates the overall quantum circuit architecture used in our work. Each qubit is initialized with a trainable data-encoding gate of the form $R_X(w_i x_i)$, where $x_i$ is the $i$-th component of the classical input vector and $w_i$ is a trainable scaling factor. This mechanism allows the quantum model to learn how to embed classical data optimally into the quantum state space.

The entangling block (shaded in blue) follows the \texttt{StronglyEntanglingLayers} design, where each qubit undergoes a sequence of $R_Z(\theta_i)$, $R_Y(\phi_i)$, and $R_Z(\delta_i)$ rotations, followed by controlled-NOT (CNOT) gates that create entanglement across qubits in a ring topology. All parameters $\theta_i$, $\phi_i$, and $\delta_i$ are independently trainable and repeated in two such layers to enhance expressivity.

Finally, a measurement in the computational (Z) basis is performed on selected qubits. The measured expectation value is post-processed by multiplying with a trainable output weight $w_\text{out}$ to yield the final continuous action used by the policy.

\subsection{In-Output Weights}
The inclusion of trainable input and output weights is inspired by the architecture proposed in~\cite{skolik2022quantum}, where classical parameters are integrated directly into quantum gate rotations. Specifically, each input $x_i$ is encoded using a trainable rotation gate $R_X(w_i x_i)$, where $w_i$ learns how strongly each feature contributes to the quantum state preparation.

After measurement, the resulting quantum expectation value is scaled by an additional trainable parameter $w_{\text{out}}$. This allows the model to learn the proper amplitude for mapping quantum measurements to continuous action values, thus avoiding the need for additional classical postprocessing.

This architecture contains all trainable parameters within the quantum circuit itself and entirely replaces the classical neural network policy in the classical MOOSE algorithm. It ensures a fully quantum policy model that directly processes classical input states and outputs continuous control actions.

\section{Infrastructures}
\subsection{Real Cart-Pole}
To further investigate the potential of quantum computing in RL, we successfully built a real-world deployment environment. 
The setup is based on a Raspberry Pi 5, which provides a Python runtime to host and execute the trained quantum policy. 
TensorFlow and TensorCircuit are used to simulate the VQCs.
The policy interacts with the physical system via an Arduino board for real-time data transmission.

Specifically, the Arduino handles motor commands and streams sensor data, including position, angle, and velocity, between the environment and the Pi.

This real-world integration enables closed-loop interaction, in which the quantum policy selects continuous actions based on real-time sensor feedback. The system operates with an average action loop of approximately 20 ms, which is sufficient for dynamic control tasks such as the cart-pole swing-up and balancing.

\subsection{MOOSE Policy}
The 
MOOSE
algorithm is designed to address the challenges of offline RL by leveraging a model-based approach to enhance sample efficiency and stability~\cite{swazinna2021overcoming}. MOOSE integrates several key components to ensure robust policy learning from a fixed dataset without additional environment interactions.

\begin{figure}[!htbp]
    \centering
    \includegraphics[width=0.95\linewidth]{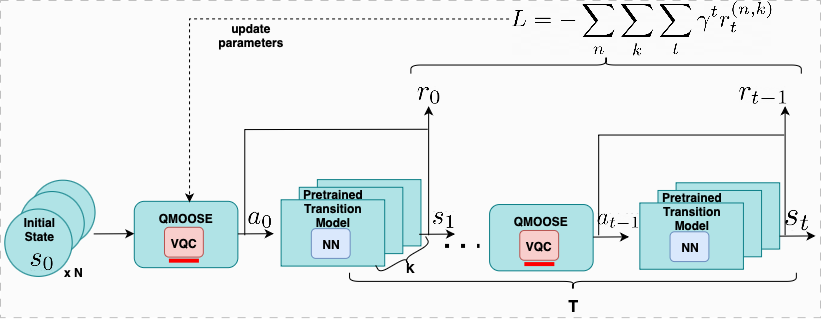}
    \caption{Rollouts of QMOOSE Policy}
    \label{fig:rollouts Qmoose}
\end{figure}
The Quantum MOOSE (QMOOSE) algorithm extends MOOSE by replacing its policy neural network with a trainable quantum circuit as shown in~\ref{fig:rollouts Qmoose}. QMOOSE comprises two main components: the transition model and the policy model. In our work, the transition model is pre-trained and kept fixed during policy learning.
It is used to predict future states from offline data. We replace the original policy model in MOOSE with our own VQC-based policy. This VQC takes the current state as input and outputs an action, interacting with the fixed transition model through model-based rollouts to guide policy optimization.
\paragraph{Transition Model}
The transition model $M$ in QMOOSE is an ensemble of neural networks trained to predict the next state given a current state-action pair. In our implementation, it is pre-trained using offline interaction data collected from the real cart-pole environment. Specifically, the dataset consists of tuples in the form of $(s_t, a_t, s_{t+1})$, where~$s_t$ denotes the current state, $a_t$ the action taken, and $s_{t+1}$ the resulting next state. A total of $K$ transition models are pre-trained independently. This ensemble approach not only captures uncertainty in the dynamics, but also mitigates model bias by aggregating predictions across multiple models. Once trained, the ensemble serves as a surrogate simulator that generates synthetic rollouts to facilitate policy improvement without requiring further real-environment interactions.

\paragraph{Policy Model}
The policy model takes as input the current state, represented by a tuple in the form of $(p_t, \dot{p_t}, \cos(\theta_t), \sin(\theta_t), \dot{\theta_t}, a_{t-3}, a_{t-2}, a_{t-1})$, where $p_t$ denotes the current cart position, $\dot{p_t}$ the cart velocity, $\theta_t$ the pole angle, $\dot{\theta_t}$ the angular velocity of the pole, and three passed actions. The inclusion of $\cos(\theta_t)$ and $\sin(\theta_t)$ ensures a continuous and unambiguous representation of the pole angle. It is trained using model-based rollouts generated by the transition model. The training performance is therefore highly dependent on the output of the transition model learned from offline data. During training, the policy model interacts closely with the transition model: it proposes actions which are then evaluated using the predicted outcomes from the learned transition dynamics. This iterative interaction enables the policy to improve based on simulated experiences, without requiring new data from the real environment.

\paragraph{Reward and Loss}
The reward function $r_t(p_t,\theta_t,\dot{\theta_t})$ is designed to encourage the cart to remain near the center and the pole to stay upright with minimal angular velocity. The reward penalizes large deviations from the center position, high pole angle, and high angular velocity. The angular velocity is conditionally applied only when the pole angle is within $-15^\circ$ to $15^\circ$. 

To account for model uncertainty and improve the robustness of policy evaluation, we use an ensemble of $K$ transition models rather than a single model. At the beginning of the training, $N$ initial states are randomly sampled from the offline dataset to serve as starting points for model-based rollouts.
The total return is obtained as the expected value of the policy's performance over model-based rollouts, defined as the cumulative reward gathered by following the policy in the learned transition dynamics. Formally, the expected return $R(\pi)$ is given by:

\begin{equation}
R(\pi) = \mathbb{E}_{s^n \sim D, \hat{M_k} \sim M}\mathbb{E}_{\tau^{(n,k)} \sim \pi, \hat{M_k}} \left[ \sum_{t}\gamma^t r^{(n,k)}_t\right],  
\end{equation}
here $\tau^{(n,k)} = (s^n_0, a_0, r_0, s^n_1, a_1, r_1, \ldots, r_T)$ denotes a trajectory sampled by executing the policy $\pi$ within the learned transition model $\hat{M}_k$. The state $s^n_0$ is the $n$-th initial position sampled from the dataset, $r_t$ is the reward at time step $t$, and $\gamma \in [0,1]$ is the discount factor.

We define the loss function as the total (negative) return, which corresponds to the cumulative cost incurred by the policy. Since rewards are negative by design, minimizing this loss encourages the agent to reduce its cumulative penalty and drive the return closer to zero:

\begin{equation}
   \mathcal{L}(\pi) = -R(\pi). 
\end{equation}
Assuming the reward function is differentiable, the gradients of this loss w.r.t. the policy parameters are computed through model-based rollouts using the fixed transition models, enabling end-to-end training of the policy.

\section{Training}
\subsection{Data Preparation}
The offline dataset for the cart-pole task was collected directly from a real physical cart-pole environment. This dataset consists of sequences of state transitions, actions, and associated rewards, capturing the dynamics observed during system operation.

\begin{figure}[!htbp]
    \centering
    \includegraphics[width=\linewidth]{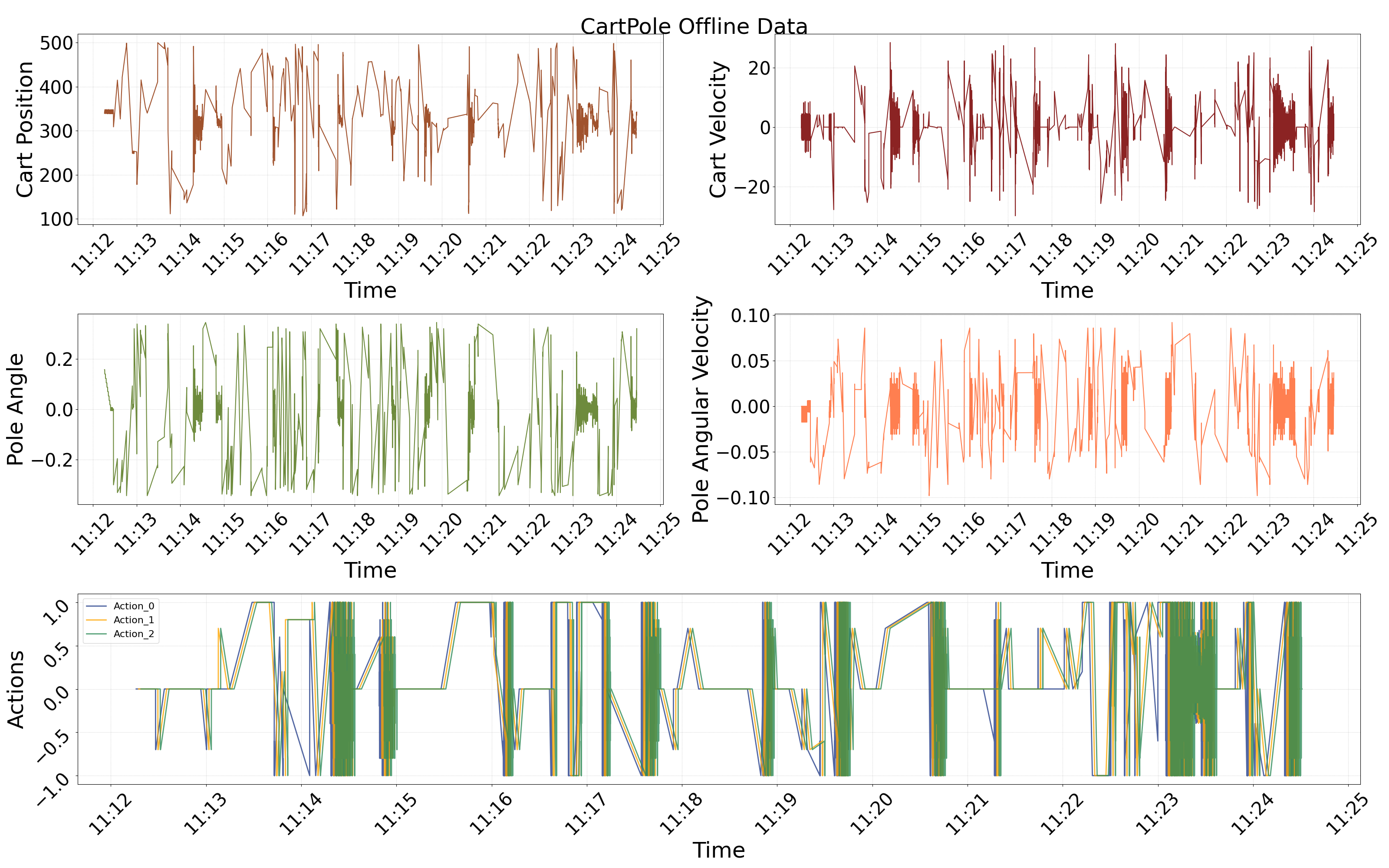}
    \caption{Offline cart-pole dataset with multiple time-series plots depicting key features such as cart position, velocity, pole angle, angular velocity, episode terminations, rewards, and actions taken.}
    \label{fig:cartpole_dataset}
\end{figure}

Figure \ref{fig:cartpole_dataset} provides an overview of the offline dataset through time-series plots. These include cart position and velocity, pole angle and angular velocity, episode termination events, reward signals, and the continuous actions applied. This variety reflects the dynamics and decision signals relevant for training RL policies.

Before training, we preprocessed the dataset to remove invalid or extreme observations, such as out-of-bounds positions or discontinuous angles. The cleaned dataset is then used to sample 400 initial states randomly. Each state comprises eight consecutive time steps with a total of 64 values, including the cart position, velocity, cosine and sine representations of the pole angle, pole angular velocity, and three previous actions.





\begin{figure}[!htbp]
    \centering
    \includegraphics[width=1.0\linewidth]{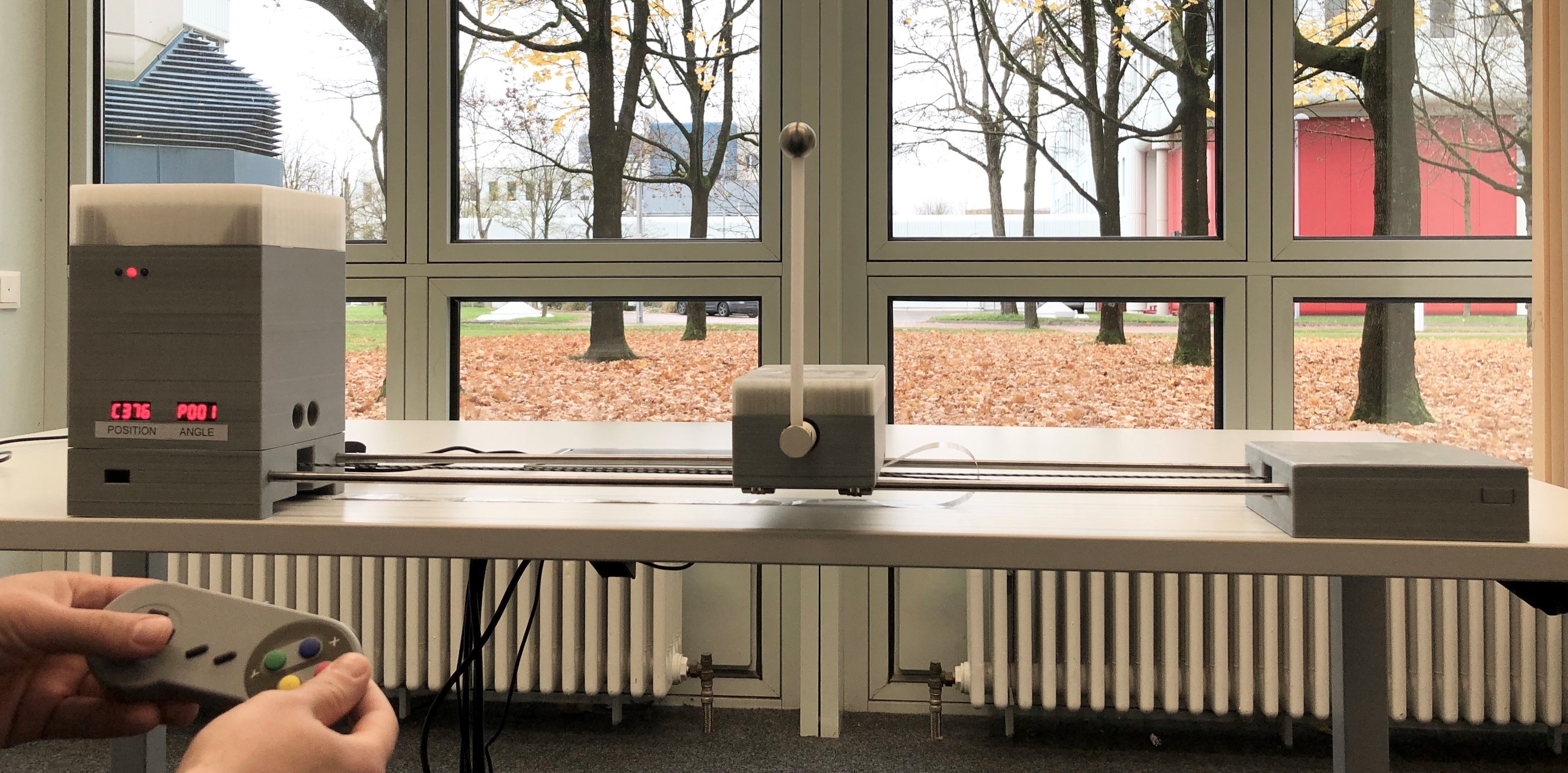}
    \caption{The real cart-pole system consists of a cart moving along a track, with a hinged pole that must be stabilized. Forces applied to the cart influence the pole’s motion. Figure adapted from \cite{hein2020interpretable}.}
    \label{fig:real_cartpole_schematic}
\end{figure}

The structured dataset allows for offline training policies in a surrogate simulation environment. After training, the policies are evaluated in the real cart-pole hardware setup (Figure~\ref{fig:real_cartpole_schematic}), where they must handle additional complexities such as sensor noise, execution latency, and mechanical imperfections.

\subsection{Hyperparameters}
Selecting optimal hyperparameters is essential for achieving good performance in RL models. We therefore performed hyperparameter optimization using Optuna \cite{akiba2019optunanextgenerationhyperparameteroptimization}, a Bayesian optimization framework that efficiently searches large hyperparameter spaces.

The key hyperparameters optimized for our VQC-based policy included the learning rate, number of variational layers, data reuploading, type of variational layers, optimizer choice, and the inclusion of trainable input and output weights.

After conducting 200 optimization trials, the final selected hyperparameters were as follows: an ensemble size of 20 transition models, Adam optimizer \cite{kingma2017adammethodstochasticoptimization} with a learning rate of 0.01, two variational layers using a strongly entangling architecture, data reuploading enabled, and both input and output weights set as trainable.

\section{Evaluation}
This section evaluates the feasibility of using a pure quantum policy model to control a \emph{real-world} physical cart-pole system. While most existing approaches in QRL are evaluated in simulation and often rely on hybrid quantum-classical architectures, our approach uses a pure VQC without any classical neural network components and evaluates it in a real-world environment.

The trained policy is deployed on a physical cart-pole setup, using a Raspberry Pi and Arduino for real-time interaction. This introduces several challenges not present during training, such as hardware latency, sensor noise, and mechanical imperfections. As such, we define a practical baseline success criterion: the ability of the quantum policy to balance the real cart-pole for at least 200 consecutive steps.

In the following subsections, we present analyses of policy performance under this setup. We compare the pure quantum policy against classical baselines and examine its stability, responsiveness, and consistency across varying initial conditions.
\begin{figure}[!htbp]
    \centering
    \includegraphics[width=\linewidth]{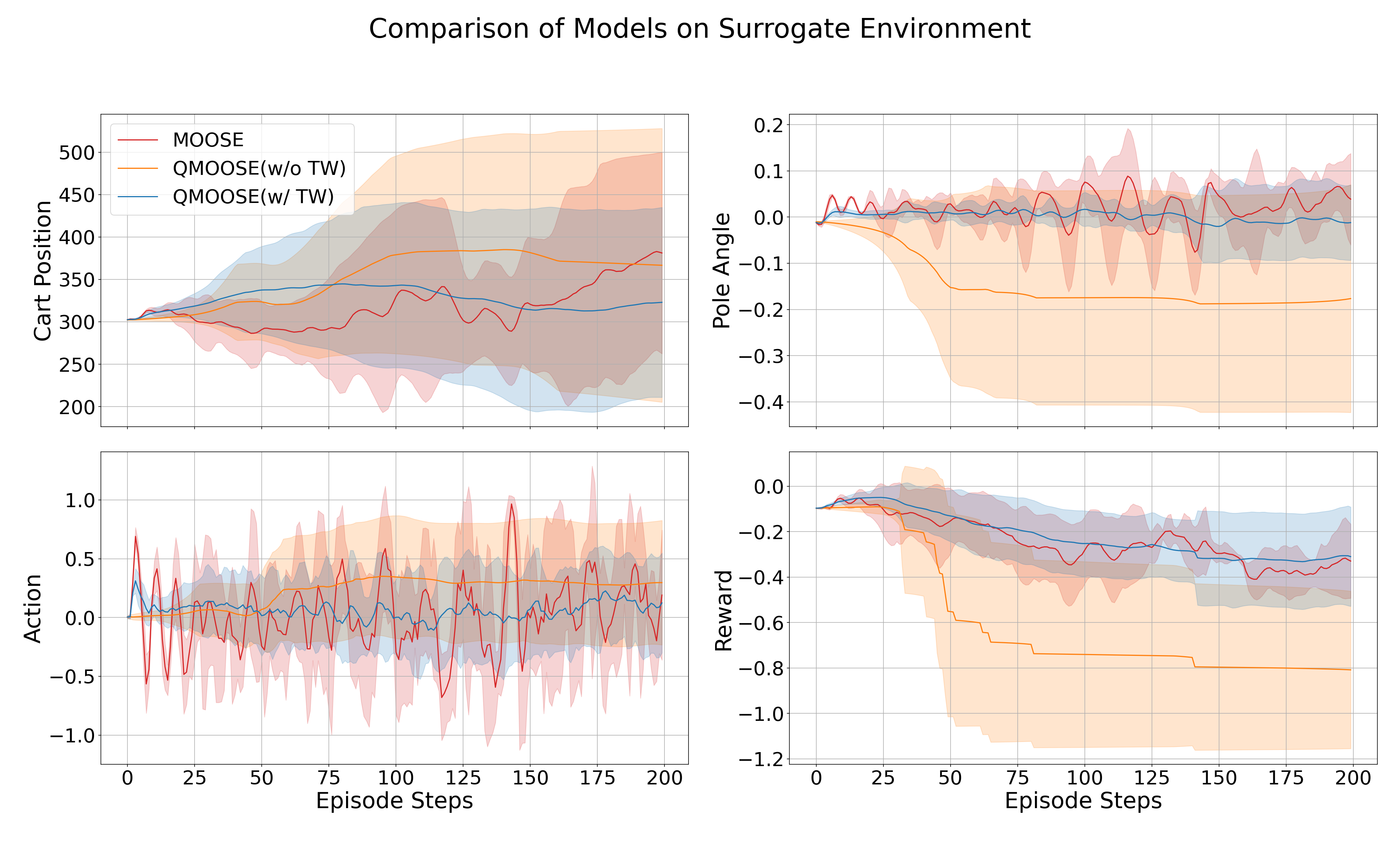}
    \caption{Comparison of models on the surrogate environment. Results are shown for classical MOOSE and QMOOSE with and without trainable input-output weights (w/ and w/o TW). Each configuration was evaluated over 5 independent runs.}
    \label{fig:ablation_moose_qmoose}
\end{figure}
\paragraph{Pre-Evaluation in Surrogate Environment}

Before deploying the learned policies on the real cart-pole hardware, we first evaluate their behavior in the surrogate environment provided by the learned transition model. This intermediate step serves as a safety check to verify that the policies produce stable and meaningful control behavior in simulation before real-world execution.

Figure~\ref{fig:ablation_moose_qmoose} compares the performance of the classical MOOSE policy and our quantum-enhanced QMOOSE policy across four key metrics: cart position, pole angle, actions, and rewards, all measured over 200 steps in the surrogate model. The shaded areas indicate standard deviation across repeated episodes.

In the surrogate environment, classical MOOSE and QMOOSE (w/ TW) are able to balance the cart-pole for the full episode length, while the QMOOSE (w/o TW) failed to make a balance. In particular, QMOOSE (w/ TW) exhibits smoother dynamics in both the pole angle and the action trajectories. MOOSE, on the other hand, produces more abrupt corrective actions, resulting in a slightly higher variance across the trajectory. While the cumulative reward trends are similar, the smoother control exhibited by QMOOSE (w/ TW) motivated us to proceed with hardware testing. However, as discussed in the next paragraph, this surrogate performance does not fully translate to robust real-world control, highlighting the simulation-to-reality gap and the challenge of policy generalization under physical noise and constraints.

\paragraph{Evaluation on Hardware}

Following the positive results in the surrogate environment, we proceeded to deploy the trained policies on the real cart-pole hardware.

To systematically evaluate the quantum policy’s robustness, we conduct a series of tests starting from a wide range of initial cart positions on the physical track. The goal is to assess whether the policy generalizes beyond specific trained regions and maintains stability across diverse conditions. We divide the track into 10-unit-wide bins and group test episodes accordingly. Each bin contains 10 evaluation runs starting from similar positions.

\begin{figure}[t]
    \centering
    {\includegraphics[width=1\linewidth]{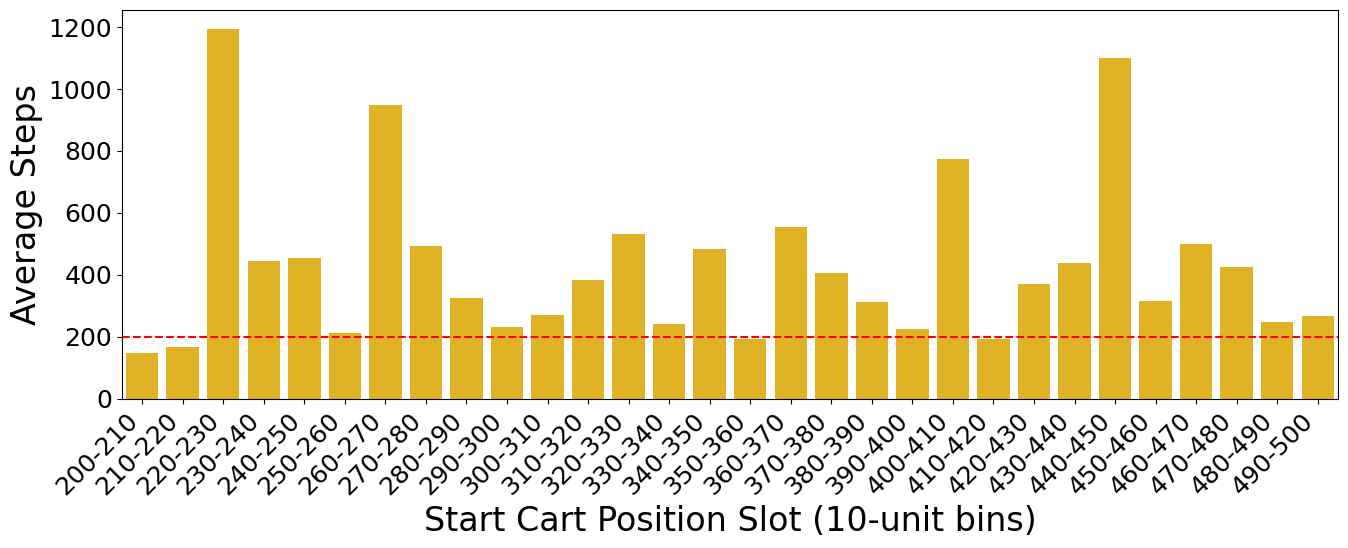}\label{fig:bar-dist}}\\[-0.05em]
    \caption{Comparison of distribution of average steps per start position slot}
    \label{fig:bar-violin}
\end{figure}

Figure~\ref{fig:bar-violin} shows the average number of steps the policy balances the cart-pole per start bin. In particular, most bins exceed the baseline threshold of 200 steps, indicating that the QMOOSE policy is capable of maintaining stability in many regions of the state space. However, a few bins near the track boundaries show reduced average steps, suggesting that the policy has difficulty recovering from edge cases or reacting to rapidly diverging initial conditions. Another possible reason is that the training data is more concentrated in the middle of the distribution than at the edges, resulting in the policy being poorly trained on less frequent input regions. These cases contribute significantly to the overall instability observed in the full evaluation. These results show that while the QMOOSE is generally able to balance the cart-pole for extended periods, its performance is not uniformly reliable across the entire state space.


\begin{figure}[!htbp]
\centering
\includegraphics[width=1\linewidth]{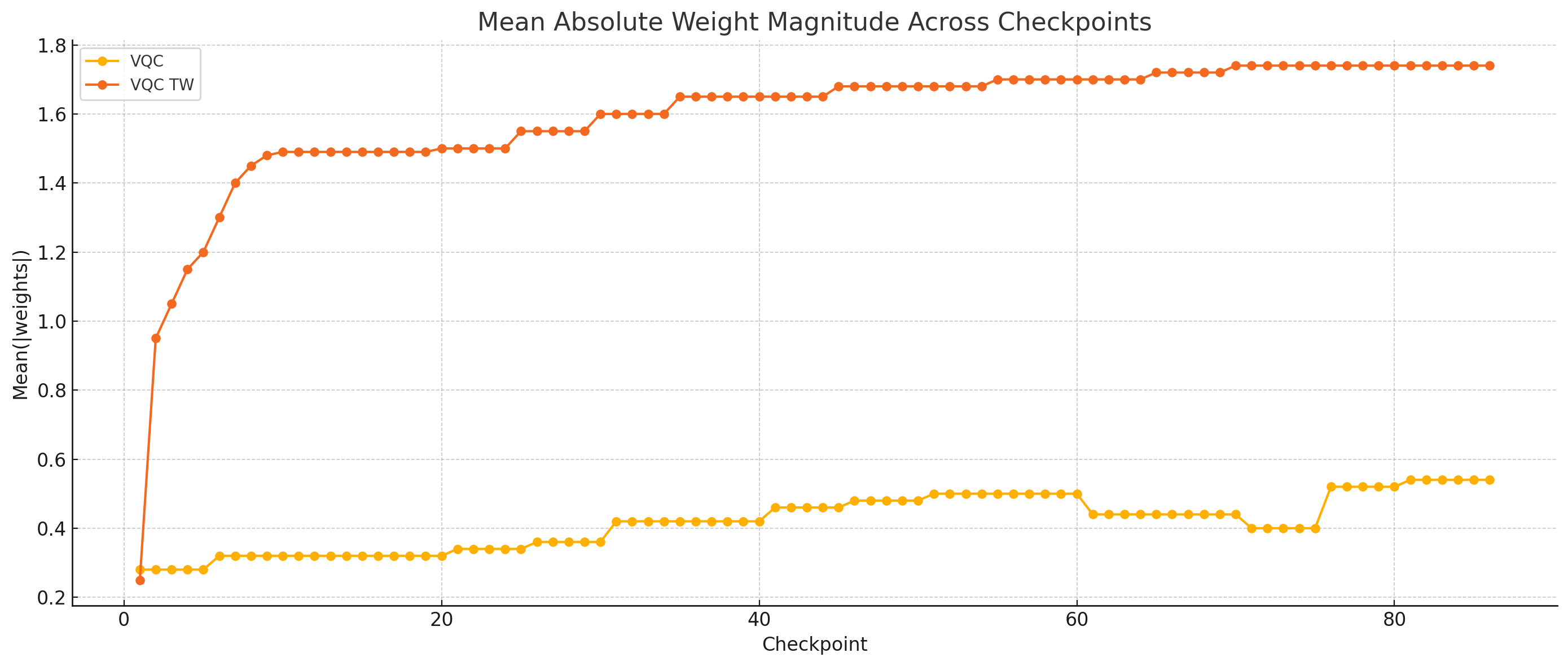}
\caption{Mean absolute weight magnitudes across checkpoints for Pure VQC and VQC with Trainable Weights.
}
\label{fig:mean_weights}
\end{figure}

Table~\ref{tab:inference-loss} shows the inference time and final training loss for the MOOSE and QMOOSE policies. Real-time control requires inference latency below 15\,ms, and both policies meet this constraint. To ensure fast execution, we selected \texttt{tensorcircuit}~\cite{zhang2023tensorcircuit} for simulating our quantum circuits, achieving less than 5\,ms inference time on standard hardware. The final training loss of QMOOSE is around $-12$, comparable to MOOSE. The QMOOSE without input-output weights achieves a training loss of approximately $-18$. Its relatively lower mean absolute weight magnitudes, as shown in Fig.~\ref{fig:mean_weights}, may partly explain its poor learning performance.

\begin{table}[t]
\centering
\caption{Comparison of Inference Time and Final Training Loss.\newline Inference time must remain below 15\,ms for real-time control. TW indicate trainalbe input and output weights}
\label{tab:inference-loss}
\begin{tabular}{lcc}
\toprule
\textbf{Policy} & \textbf{Inference Time (ms)} & \textbf{Final Training Loss} \\
\midrule
MOOSE (Classical) & 1.5 $\pm$ 0.5 & $-10.2$ \\
QMOOSE (w/o TW) & - & $-18.0$\\
QMOOSE (w/ TW) & 4 $\pm$ 1 & $-12.5$ \\
\bottomrule
\end{tabular}
\end{table}

Figure~\ref{fig:mean_weights} illustrates the mean absolute weight magnitudes across training checkpoints, comparing the VQC model to our proposed approach using VQC with trainable input and output weights (VQC TW).
The “VQC" model uses the same circuit structure as our proposed approach, but omits trainable input and output weights, relying instead on fixed encodings and unscaled measurement outputs.

Table~\ref{tab:QPU vs CPU} presents a breakdown of the time cost of executing a single QMOOSE policy action. On the Raspberry Pi 5, all components including CPU-based preprocessing, circuit simulation, and postprocessing are executed locally, with a time cost of approximately $5$ ms, satisfying real-time control requirements. In contrast, the IBM Cloud-based execution introduces significant overhead: although the CPU component is relatively fast, cloud communication alone adds roughly $700$ ms, and the quantum circuit execution on the QPU takes an additional $3000$ ms. Queue time is highly variable and thus not included in this analysis. These measurements highlight a major bottleneck in cloud-based quantum control systems, with overall latency exceeding $3.7$ seconds per action, and underscore the need to rely on simulated quantum circuits for latency-sensitive applications.

\begin{table}[t]
\centering
\caption{Latency breakdown for executing a single QMOOSE action.}
\label{tab:QPU vs CPU}
\begin{tabular}{lcc}
\toprule
 & \textbf{Raspberry Pi (ms)} & \textbf{IBM Cloud QPU (ms)} \\
\midrule
CPU & 5 & 1 \\
Cloud Delay & -- & 700 \\
Queue Time & -- & -- \\
QPU & -- & 3000 \\
\bottomrule
\end{tabular}
\begin{tablenotes}
\small
\item \textbf{Note:} The Raspberry Pi setup runs all components locally using a simulated backend. The IBM Cloud QPU setup includes local preprocessing, remote quantum circuit execution, and network delays. Queue time is omitted due to variability.
\end{tablenotes}
\end{table}

\section{Conclusion, Limitations and Future Directions}
\subsection{Conclusion}


This work investigated the use of quantum-enhanced policies in offline RL by extending the MOOSE algorithm.
Specifically, we replaced the classical multilayer perceptron with a VQC and introduced trainable input-output weights. While the pure VQC model struggled with learning, the addition of trainable weights improved learning stability and led to a partially successful control of a real-world cart-pole system. However, real-time deployment on cloud-based quantum hardware remains infeasible due to high latency, whereas local simulation on a Raspberry Pi 5 meets real-time constraints.

\subsection{Limitations}
Despite promising results, our study faced several limitations. The offline RL dataset restricted generalization, as it did not cover all real-world state-action pairs. Our low-cost cart-pole setup also introduced imperfections such as latency. Finally, since our focus was on algorithmic benchmarking, we relied on simulated quantum hardware to ensure reproducibility and isolate performance from noise; evaluating on real quantum devices remains future work.

\subsection{Future Directions}
While our results demonstrate that quantum-enhanced policies could be successfully deployed in real-world control tasks, several avenues for future research remain open.


\paragraph{Enhanced Quantum Circuit Architectures}
More expressive circuits may improve performance. Automated search methods such as Differentiable Quantum Architecture Search \cite{sun2024differentiable, chen2024differentiablequantumarchitecturesearch} or deeper entanglement structures \cite{Eisenmann_2024} could help.




\paragraph{Alternative Quantum Encoding Schemes}
We used $R_X$-based encoding; future studies could explore alternatives such as amplitude encoding \cite{schuld2018supervised} or tensor network-based representations \cite{perezgarcia2007matrixproductstaterepresentations} for better scalability in high-dimensional spaces.

\paragraph{Hybrid Quantum Framework}
An important future direction lies in the development of hybrid quantum-classical architectures that can mitigate the impact of execution delay. By batched inference, we could evaluate quantum policies over batches of states and store likely actions for future states.


\printbibliography

@article{skolik2022quantum,
  title         = {Quantum Agents in the {Gym}: A Variational Quantum Algorithm for Deep {Q-learning}},
  author={Skolik, Andrea and Jerbi, Sofiene and Dunjko, Vedran},
  journal={Quantum},
  volume={6},
  pages={720},
  year={2022},
  doi           = {10.22331/q-2022-05-24-720},
  publisher={Verein zur F{\"o}rderung des Open Access Publizierens in den Quantenwissenschaften}
}

@article{hein2020interpretable,
  title={Interpretable control by reinforcement learning},
  author={Hein, Daniel and Limmer, Steffen and Runkler, Thomas A},
  journal={IFAC-PapersOnLine},
  volume={53},
  number={2},
  pages={8082--8089},
  year={2020},
  publisher={Elsevier}
}

@inproceedings{sun2024differentiable,
  title={Differentiable quantum architecture search for job shop scheduling problem},
  author={Sun, Yize and Liu, Jiarui and Ma, Yunpu and Tresp, Volker},
  booktitle={ICASSP 2024-2024 IEEE International Conference on Acoustics, Speech and Signal Processing (ICASSP)},
  pages={236--240},
  year={2024},
  organization={IEEE}
}

@inproceedings{kolle2024study,
  title={A study on optimization techniques for variational quantum circuits in reinforcement learning},
  author={K{\"o}lle, Michael and Witter, Timo and Rohe, Tobias and Stenzel, Gerhard and Altmann, Philipp and Gabor, Thomas},
  booktitle={2024 IEEE International Conference on Quantum Software (QSW)},
  pages={157--167},
  year={2024},
  organization={IEEE}
}

@article{zhang2023tensorcircuit,
  title={Tensorcircuit: a quantum software framework for the nisq era},
  author={Zhang, Shi-Xin and Allcock, Jonathan and Wan, Zhou-Quan and Liu, Shuo and Sun, Jiace and Yu, Hao and Yang, Xing-Han and Qiu, Jiezhong and Ye, Zhaofeng and Chen, Yu-Qin and others},
  journal={Quantum},
  volume={7},
  pages={912},
  year={2023},
  publisher={Verein zur F{\"o}rderung des Open Access Publizierens in den Quantenwissenschaften}
}

@article{swazinna2021overcoming,
  title={Overcoming model bias for robust offline deep reinforcement learning},
  author={Swazinna, Phillip and Udluft, Steffen and Runkler, Thomas},
  journal={Engineering Applications of Artificial Intelligence},
  volume={104},
  pages={104366},
  year={2021},
  publisher={Elsevier}
}

@article{Bennett_2014,
   title={Quantum cryptography: Public key distribution and coin tossing},
   volume={560},
   ISSN={0304-3975},
   url={http://dx.doi.org/10.1016/j.tcs.2014.05.025},
   DOI={10.1016/j.tcs.2014.05.025},
   journal={Theoretical Computer Science},
   publisher={Elsevier BV},
   author={Bennett, Charles H. and Brassard, Gilles},
   year={2014},
   month=dec, pages={7–11} 
}

@article{aspuru2005simulated,
  title={Simulated quantum computation of molecular energies},
  author={Aspuru-Guzik, Al{\'a}n and Dutoi, Anthony D and Love, Peter J and Head-Gordon, Martin},
  journal={Science},
  volume={309},
  number={5741},
  pages={1704--1707},
  year={2005},
  publisher={American Association for the Advancement of Science}
}

@article{farhi2019quantumsupremacyquantumapproximate,
      title={Quantum Supremacy through the Quantum Approximate Optimization Algorithm}, 
      author={Edward Farhi and Aram W Harrow},
      year={2019},
      journal={arXiv preprint},
      doi={10.48550/arXiv.1602.07674},
}

@article{Cerezo_2021,
   title={Cost function dependent barren plateaus in shallow parametrized quantum circuits},
   volume={12},
   ISSN={2041-1723},
   url={http://dx.doi.org/10.1038/s41467-021-21728-w},
   DOI={10.1038/s41467-021-21728-w},
   number={1},
   journal={Nature Communications},
   publisher={Springer Science and Business Media LLC},
   author={Cerezo, M. and Sone, Akira and Volkoff, Tyler and Cincio, Lukasz and Coles, Patrick J.},
   year={2021},
   month=mar 
}

@article{Kober_2013,
author = {Kober, Jens and Bagnell, J. and Peters, Jan},
year = {2013},
month = {09},
pages = {1238-1274},
title = {Reinforcement Learning in Robotics: A Survey},
volume = {32},
isbn = {978-3-642-27644-6},
journal = {The International Journal of Robotics Research},
doi = {10.1177/0278364913495721}
}

@misc{yu2020reinforcementlearninghealthcaresurvey,
      title={Reinforcement Learning in Healthcare: A Survey}, 
      author={Chao Yu and Jiming Liu and Shamim Nemati},
      year={2020},
      eprint={1908.08796},
      archivePrefix={arXiv},
      primaryClass={cs.LG},
      url={https://arxiv.org/abs/1908.08796}, 
}

@article{LI202375,
title = {Deep reinforcement learning in smart manufacturing: A review and prospects},
journal = {CIRP Journal of Manufacturing Science and Technology},
volume = {40},
pages = {75-101},
year = {2023},
issn = {1755-5817},
doi = {https://doi.org/10.1016/j.cirpj.2022.11.003},
url = {https://www.sciencedirect.com/science/article/pii/S1755581722001717},
author = {Chengxi Li and Pai Zheng and Yue Yin and Baicun Wang and Lihui Wang}
}

@article{Dunjko_2016,
   title={Quantum-Enhanced Machine Learning},
   volume={117},
   ISSN={1079-7114},
   url={http://dx.doi.org/10.1103/PhysRevLett.117.130501},
   DOI={10.1103/physrevlett.117.130501},
   number={13},
   journal={Physical Review Letters},
   publisher={American Physical Society (APS)},
   author={Dunjko, Vedran and Taylor, Jacob M. and Briegel, Hans J.},
   year={2016},
   month=sep 
}

@misc{chen2020variationalquantumcircuitsdeep,
      title={Variational Quantum Circuits for Deep Reinforcement Learning}, 
      author={Samuel Yen-Chi Chen and Chao-Han Huck Yang and Jun Qi and Pin-Yu Chen and Xiaoli Ma and Hsi-Sheng Goan},
      year={2020},
      eprint={1907.00397},
      archivePrefix={arXiv},
      primaryClass={cs.LG},
      url={https://arxiv.org/abs/1907.00397}, 
}

@misc{kölle2024quantumadvantageactorcriticreinforcement,
      title={Quantum Advantage Actor-Critic for Reinforcement Learning}, 
      author={Michael Kölle and Mohamad Hagog and Fabian Ritz and Philipp Altmann and Maximilian Zorn and Jonas Stein and Claudia Linnhoff-Popien},
      year={2024},
      eprint={2401.07043},
      archivePrefix={arXiv},
      primaryClass={quant-ph},
      url={https://arxiv.org/abs/2401.07043}, 
}

@misc{akiba2019optunanextgenerationhyperparameteroptimization,
      title={Optuna: A Next-generation Hyperparameter Optimization Framework}, 
      author={Takuya Akiba and Shotaro Sano and Toshihiko Yanase and Takeru Ohta and Masanori Koyama},
      year={2019},
      eprint={1907.10902},
      archivePrefix={arXiv},
      primaryClass={cs.LG},
      url={https://arxiv.org/abs/1907.10902}, 
}

@inproceedings{Eisenmann_2024,
   title={Model-Based Offline Quantum Reinforcement Learning},
   url={http://dx.doi.org/10.1109/QCE60285.2024.00175},
   DOI={10.1109/qce60285.2024.00175},
   booktitle={2024 IEEE International Conference on Quantum Computing and Engineering (QCE)},
   publisher={IEEE},
   author={Eisenmann, Simon and Hein, Daniel and Udluft, Steffen and Runkler, Thomas A.},
   year={2024},
   month=sep, pages={1490–1496} 
}

@misc{chen2024differentiablequantumarchitecturesearch,
      title={Differentiable Quantum Architecture Search in Asynchronous Quantum Reinforcement Learning}, 
      author={Samuel Yen-Chi Chen},
      year={2024},
      eprint={2407.18202},
      archivePrefix={arXiv},
      primaryClass={quant-ph},
      url={https://arxiv.org/abs/2407.18202}, 
}

@book{schuld2018supervised,
author = {Schuld, Maria and Petruccione, Francesco},
title = {Supervised Learning with Quantum Computers},
year = {2018},
isbn = {3319964232},
publisher = {Springer Publishing Company, Incorporated},
edition = {1st},
doi={10.1007/978-3-319-96424-9}
}

@article{perezgarcia2007matrixproductstaterepresentations,
      title={Matrix Product State Representations}, 
      author={D. Perez-Garcia and F. Verstraete and M. M. Wolf and J. I. Cirac},
      year={2007},
      journal={arXiv preprint},
      doi={10.48550/arXiv.0608197},
}

@misc{kingma2017adammethodstochasticoptimization,
      title={Adam: A Method for Stochastic Optimization}, 
      author={Diederik P. Kingma and Jimmy Ba},
      year={2017},
      eprint={1412.6980},
      archivePrefix={arXiv},
      primaryClass={cs.LG},
      url={https://arxiv.org/abs/1412.6980}, 
}

@article{Schuld_2020,
   title={Circuit-centric quantum classifiers},
   volume={101},
   ISSN={2469-9934},
   url={http://dx.doi.org/10.1103/PhysRevA.101.032308},
   DOI={10.1103/physreva.101.032308},
   number={3},
   journal={Physical Review A},
   publisher={American Physical Society (APS)},
   author={Schuld, Maria and Bocharov, Alex and Svore, Krysta M. and Wiebe, Nathan},
   year={2020},
}

\end{document}